\documentclass[smallcondensed]{svjour2}

\smartqed  
\usepackage{color}
\usepackage{graphicx}
\usepackage{amsmath}
\usepackage{amssymb}
\usepackage{times}
\usepackage{hyperref} 
\usepackage{epsfig}
\usepackage{psfrag}

\newcommand{\eqnsection}{
\renewcommand{\theequation}{\thesection.\arabic{equation}}
    \makeatletter
    \csname  @addtoreset\endcsname{equation}{section}
    \makeatother}
\eqnsection

\def\P{{\bf P}}
\def\E{{\bf E}}

\usepackage[english]{babel}
\usepackage[latin1]{inputenc}
\usepackage{color,a4,pstricks}
\usepackage[T1]{fontenc}
\usepackage{graphicx}
\usepackage{dcolumn}
\usepackage{bm}
\usepackage{amsmath}
\journalname{J Stat Phys}
\begin{document}
\title{
 Large deviations for the branching Brownian motion in presence of  selection or    coalescence
}

\author{Bernard Derrida \and Zhan Shi}

\institute{   B. Derrida  \\
Coll\`ege de France, 11 place Marcelin Berthelot, 75231 Paris Cedex 05 - France\\
and  \\
              Laboratoire de Physique Statistique,\\
              \'Ecole Normale Sup\'erieure,
 Universit\'e Pierre et Marie Curie, Universit\'e Denis Diderot, CNRS \\
              24 rue Lhomond,
              75231 Paris Cedex 05 - France \\
\\ \ \\   Z. Shi  \\ 
LPMA, Universit\'e Pierre et Marie Curie \\
4 place Jussieu, 75252 Paris Cedex 05 - France 
           \\   \email{derrida@lps.ens.fr}\and\email{zhan.shi@upmc.fr}
}

\date{Received: date / Accepted: date}
\maketitle

\begin{abstract}
The large deviation function has been known for a long time in the literature for the displacement of the rightmost particle in a branching random walk (BRW), or in a branching Brownian motion (BBM).
More recently a number of generalizations of the  BBM and of the BRW have been considered where  selection or  coalescence mechanisms tend to limit the exponential growth of the number of particles. Here we try to estimate the large deviation function of the position of the rightmost particle for several such generalizations: the $L$-BBM, the $N$-BBM, and the CBRW (coalescing branching random walk) which is closely related to the noisy FKPP equation.
Our approach allows us to obtain only upper bounds on these large deviation functions. One noticeable feature of our results is their non analytic dependence   on the parameters (such as the coalescence rate in the CBRW).

\end{abstract}
\ \\ \ \\
\today \\ \ \\

\section{Introduction}
 Branching Brownian motions (BBM) and branching random walks (BRW)  are among the simplest stochastic models of a growing population in space and time.
They describe particles which perform Brownian motions or random walks and branch independently at random times \cite{berestycki,stf,zeitouni}.
If one starts with a single particle,  the  size of the region of space occupied by the particles  grows linearly with time.
Since the mid seventies, one has a precise understanding of the fluctuations of the size of this region 
\cite{McKean,Bramson1,Bramson2,Majumdar}.
 For example in the one dimensional case one knows   that the probability distribution of the  position of the rightmost particle  of a BBM can be obtained by solving 
an FKPP (Fisher-Kolmogorov-Petrovskii-Piskounov)
 equation  \cite{McKean,Bramson1,Bramson2,vS}:  for a BBM starting at the origin,  where particles diffuse according to $$\langle [X(t) - X(0)]^2  \rangle = \sigma^2 t$$  and branch at rate $1$, one  can 
show \cite{McKean} that, at time $t$,
 the probability $P(x,t)$ that  the rightmost particle is on  the right of $x$  is the solution  of  the FKPP equation
\begin{equation}
{\partial P(x,t) \over \partial t} = {\sigma^2 \over 2} 
{\partial^2 P(x,t) \over \partial x^2} +P(x,t)-P^2(x,t)
\label{KPP}
\end{equation}
with  a step initial condition $P(x,0)=1-\theta(x)$ (where $\theta(x)$ is the Heaviside function).
In the long time limit,    it is known  \cite{Bramson1,Bramson2} that 
the probability $-\partial P(x,t) / \partial x$   that the position 
of the rightmost particle
 $X_{\rm max}(t)=x$  
 is concentrated around  $X_t \simeq \sqrt{2} \sigma t -{3 \sigma \over 2 \sqrt{2}} \ln t $.

One can also show \cite{CR,Rouault} from  (\ref{KPP}) that  the large deviation function $\psi_{\rm BBM}$ of the position $X_{\rm max}(t)$  of the rightmost particle  for
 $v >  \sqrt{2} \sigma$ 
\begin{equation}
\label{ld-BBM}
\P (X_{\rm max} (t) \ge v t )  \sim \exp[- t \  \psi_{\rm BBM} (v) ]
\end{equation}
is given by 
\begin{equation}
   \psi_{\rm BBM} (v) = {v^2 \over 2 \sigma^2}-1 \ .  \label{psi-BBM} \end{equation}
 In (\ref{ld-BBM})  and everywhere below the symbol $\sim$  means that
\begin{equation}
\label{symbol}
\lim_{t \to \infty} {\ln 
\P (X_{\rm max} (t) \ge  v t )   \over t} = - \psi_{\rm BBM}(v) \ .  
\end{equation}

 Over the last decade a number of generalizations of the branching Brownian motion have been considered where, 
 due to some selection or coalescence mechanism, 
 the density of particles generated by the BBM saturates.
These extensions of the BBM are expected to  be described by  noisy  versions of the FKPP equation \cite{MulS,DMS}.
In these noisy versions, the  main effect of the noise is to shift the velocity of the front \cite{BD1,PL,DR,MMQ,BG,CD,Pain} and to make  its position fluctuate \cite{BD2,Panja,BDMM3}. A phenomenological approach has been proposed in \cite{BDMM3,BD2} which gives a  prediction for    the cumulants of this position. Our goal here  is  to understand the large {\it positive} deviations of this position. The case of large {\it negative} deviations   (studied in \cite{MS,MVS} for branching random walks with coalescence)  would require  a rather different approach and will not be  discussed in this paper  except for some comments  in the conclusion; in particular the large deviation function  may depend on the number of particles one starts with.

In the present work we try to study how (\ref{psi-BBM}) is modified by these  selection or coalescence
mechanisms.  We discuss three models:
\begin{enumerate}
\item {\bf The $L$-BBM \cite{BDMM3,Pain}:}

In the $L$-BBM, one starts at time $t=0$ with a single particle at the origin. This
 particle branches and diffuses like a  usual branching Brownian motion. The only difference  with the usual BBM is that whenever a particle gets at a distance larger than $L$ from the rightmost particle, it is eliminated.
Therefore 
 at any given  time $t$ the system consists of a random number ${\cal N}(t) \ge 1$ of particles at positions $X_1(t), X_2(t), \cdots X_{{\cal N}(t)}$ which all satisfy 
$$ X_{\rm max}(t) -L\le X_i(t) \le X_{\rm max}(t) $$
 where $
 X_{\rm max}(t)
=  \max_{1 \le i \le {\cal N}(t)} X_i(t) $.

This number of particles ${\cal N}(t)$ fluctuates but one can show (see the discussion in Section \ref{existence}) that the evolution of the $L$-BBM leads to a steady state where   the event  ${\cal N}(t)=1$ is recurrent.

For large $t$  one can also show (see Section \ref{existence}) that the probability distribution of   the position $X_{\rm max}(t)$ of the  rightmost particle
has a large deviation form
\begin{equation}
\P_\mathrm{LBBM} (X_{\rm max}(t)\ge v t) \sim 
 \exp[-t \  \psi_{\rm LBBM} (v) ] \ .
\label{ld-LBBM}
\end{equation}
One of our results (see Sections \ref{physical} and \ref{proofs}) is  the following    upper bound    for $v > \sqrt{2} \sigma$
and large $L$
\begin{equation}
 0 \le    \psi_{\rm LBBM} (v)  
   - \psi_{\rm BBM} (v) \lesssim e^{-  \alpha(v) \,  L / \sigma } 
\label{Res-LBBM}
\end{equation}
  with 
\begin{equation}
  \alpha(v)= \begin{cases}
  {2\sqrt{2}\, (v -v_c ) \over v_c}  & {\rm for } \ \ \ v_c  < v < {3    \over  2 }  v_c
\\ &
\\
   {v \, 
 + \,\sqrt{v^2 - 2 v_c^2}
\over \sqrt{2}\, v_c } 
  & {\rm for }  \ \ \  v > 
{3    \over  2 }  v_c
   \end{cases}
\label{Res2-LBBM} 
\end{equation}
where
\begin{equation}
v_c=\sqrt{2} \sigma
\ . 
\label{vc-eq}
\end{equation}
In (\ref{Res-LBBM})  and everywhere else in this paper, the symbol $\lesssim$ means  that 
$$\limsup_{L\to \infty}{
 \ln(   \psi_{\rm LBBM} (v)  
   - \psi_{\rm BBM} (v))\over
L/\sigma} \le -  \alpha(v) \ .  $$
\item {\bf The $N$-BBM \cite{BDMM1,BDMM2,DR,M1,BBMM,Mallein}:} \\
In the $N$-BBM one starts as above with a single particle at $t=0$ which diffuses and branches but the size of the population cannot exceed  a fixed value $N$. As long as  the number of particles ${\cal N}(t)$ is less than $N$  the evolution is exactly the same as for the BBM. 
However,  when ${\cal N}(t)=N$,  as soon as a new branching event occurs, the leftmost particle is eliminated  so that  the total number of particles  remains subsequently equal to $N$.

For the $N$-BBM we will obtain 
(see Sections \ref{physical} and \ref{proofs} below) 
 for the large deviation function 
\begin{equation}
\P_\mathrm{NBBM} (X_{\rm max}(t)\ge v t) \sim 
 \exp[-t \  \psi_{\rm NBBM} (v) ] 
\label{ld-NBBM}
\end{equation}
 an upper bound
\begin{equation}
 0 \le    \psi_{\rm NBBM} (v)  
   - \psi_{\rm BBM} (v) \lesssim N^{-  \beta(v)   } 
\label{Res-NBBM}
\end{equation}
where
\begin{equation}
  \beta(v)= \begin{cases}
{\frac{v^2}{v_c^2}-1}  & {\rm for }  \ \ \ v_c  < v \le
\sqrt{2} v_c 
\\ & \\
  {v^2  \over 2 v_c^2}  & {\rm for }  \ \ \  v \ge
\sqrt{2} v_c 
   \end{cases}
\label{Res1-NBBM} \ ,
\end{equation}
where $v_c$ is given by (\ref{vc-eq}).
In fact, as discussed in the conclusion, we believe that $\beta(v)= \frac{v^2}{v_c^2}$ remains valid even for $v> \sqrt{2} v_c$. This would follow from a conjecture (\ref{conjecture}) that we formulate in the conclusion but that we did not succeed to prove.

\

\item {\bf The CBRW (coalescing branching random walk)}
\cite{DMS,MS,MVS}

An important motivation in the study of the CBRW is its dual relation with the noised FKPP equation, rigorously established in \cite{DMS}. 

To explain how the CBRW is defined let us first consider a branching random walk BRW on a one dimensional lattice with lattice spacing $\sigma$:  a particle on site $x$ jumps to site $x + \sigma$ at rate $1/2$, to site $x-\sigma$ at rate $1/2$ and branches at rate $r$ to give rise to two new particles on the same site.

 The trajectory of each particle is a random walk and in the long time limit the probability that such a random walk reaches a position $x= v t$ is of the form
\begin{equation}
    \P_{\rm RW}(x=vt ) \sim e^{-t f(v)}
    \label{fv0}
\end{equation}

\noindent where  
\begin{equation}
 \label{fv}
f(v) =
   1 -  \sqrt{ 1+{v^2 \over  \sigma^2}} +{v \over \sigma} \ln \left({v\over  \sigma} + \sqrt{1 + {v^2 \over  \sigma^2}} \right) \ .   \end{equation}
Using the fact that $\langle e^{\lambda x} \rangle = e^{t g(\lambda)}$
with 
\begin{equation}
g(\lambda) = \cosh(\lambda \sigma) -1 ,
\label{g} \end{equation}
the large deviation function (\ref{fv}) can be easily obtained  from the  parametric form as
\begin{equation}
   f (v) =-g(\lambda) + \lambda g'(\lambda) \ \ \ \ ; \ \ \ \ v= g'(\lambda)  \ . 
\label{parametric}
\end{equation}

As  the particles branch at rate $r$, the distribution of the position
$X_{\rm max} (t) $
 of the rightmost particle
of this  BRW, (in absence of coalescence), is of the form \cite{Biggins}
\begin{equation}
\label{ld-BRW}
\P_\mathrm{BRW} (X_{\rm max} (t) \ge v t )  \sim \exp[- t \  \psi_{\rm BRW} (v) ]
\end{equation}
with 
\begin{equation}
   \psi_{\rm BRW} (v) =  f(v)-r\ .  \label{psi-BRW} 
\end{equation}

Now in the coalescing branching random walk (CBRW),   in addition to the diffusion  and the branching,  we  let each pair of particles on the same site coalesce at rate $\mu$. We will show  in Section \ref{existence}
\begin{equation}
\label{ld-CBRW}
\P_\mathrm{CBRW} (X_{\rm max} (t) \ge  v t )  \sim \exp[- t \  \psi_{\rm CBRW} (v) ]
\end{equation}
and in Sections \ref{physical} and \ref{proofs} that for $\mu \to 0$, 
\begin{equation}
 0 \le    \psi_{\rm CBRW} (v)  
   - \psi_{\rm BRW} (v) \lesssim \mu^{  \gamma(v)   }  \ , 
\label{Res-CBRW}
\end{equation}
  where
\begin{equation}
  \gamma(v)= \begin{cases}
 {f'(v) \over f'(y)} -1 \
    & {\rm for }  \ \ \ v_c < v < v_1 
\\
  1  & {\rm for }  \ \ \  v > v_1
   \end{cases}
\label{Res1-CBRW}
\end{equation}
and, where for each $v$, $y$ is solution of   
\begin{equation}
{f(y)-r\over f'(y)} - y =
{f(v)-r\over f'(v)} - v 
\label{v}
\end{equation}
with 
$v_c$ and $v_1$ given  by
\begin{equation}
 \psi_{\rm BRW}(v_c)=0 \ \ \ ; \ \ \  \gamma(v_1)=1  \ ,  
\label{vc-v1}
\end{equation}
(i.e. $v_1$ is the value of $v$ such that  $f'(v)= 2 f'(y)$).
\ \\ \ \\

The general expression (\ref{Res1-CBRW}) simplifies 
when $r \ll 1$. One then has $v_c \simeq \sqrt{2 r} \sigma$ and in the whole range $ v_c < v \ll \sigma$
$$ f(v) \simeq {v^2\over 2  \sigma^2}$$
instead of  (\ref{g}). All the other steps remain the same with $y=2 \sigma^2/v$,
$v_1= \sqrt{2} v_c$ and therefore
\begin{equation}
   \psi_{\rm BRW} (v) = {v^2 \over  v_c^2} -1
\ \ \ \ ; 
\ \ \ \  
  \gamma(v)= \begin{cases}
 {v^2 \over v_c^2} -1 \
    & {\rm for }  \ \ \ v_c < v <  \sqrt{2} v_c
\\
  1  & {\rm for }  \ \ \  v >  \sqrt{2} v_c \ . 
   \end{cases}
   \label{1.22}
\end{equation}

If one would consider more general branching  random walks,
characterized by the rate $\rho(y)$ at which a particle jumps  a distance $y$ from the site it occupies,   $g(\lambda)$ would be given by
\begin{equation}
g(\lambda)= \sum_y \rho(y) (e^{\lambda y} -1) 
\label{gnew}
\end{equation}
 and all the rest (\ref{parametric}-\ref{vc-v1})  would remain unchanged 
with only (\ref{g}) replaced by (\ref{gnew}).
\ \\ \ \\ {\it Remark:} 
A way of looking for  a solution $y$ of (\ref{v}) is to work with the Legendre transform $g(\lambda)$ related to $f(v)$ by  (\ref{parametric}).
If $\lambda_0$ and $\lambda_1$ are defined by $v=g'(\lambda_0)$ and $y=g'(\lambda_1)$, one can  check that (\ref{v}) becomes 
$${g(\lambda_1) + r \over \lambda_1}
={g(\lambda_0) + r \over \lambda_0}. $$
Under this form, one can show using the convexity of $g(\lambda)$ that $(g(\lambda)+r)/\lambda$ has a single minimum at some value $\lambda_c$, that $v_c=g'(\lambda_c)$ and that as $(g(\lambda)+r) /\lambda \to \infty$ as $\lambda \to 0$ or $\infty$ (we restrict our discussions here and below to $g(\lambda) \to \infty$ as $\lambda\to \pm \infty$),
there is always a solution $\lambda_1$ and therefore  a solution $y$ of (\ref{v}).
\\ \ \\
\end{enumerate}

\section{The physical picture}
\label{physical}
In this section we explain a heuristic way of understanding the claims in the introduction.
The main idea is rather similar for the three problems (see Figure 1).

\begin{figure}[b]
\centering{\includegraphics[width=.75\textwidth]{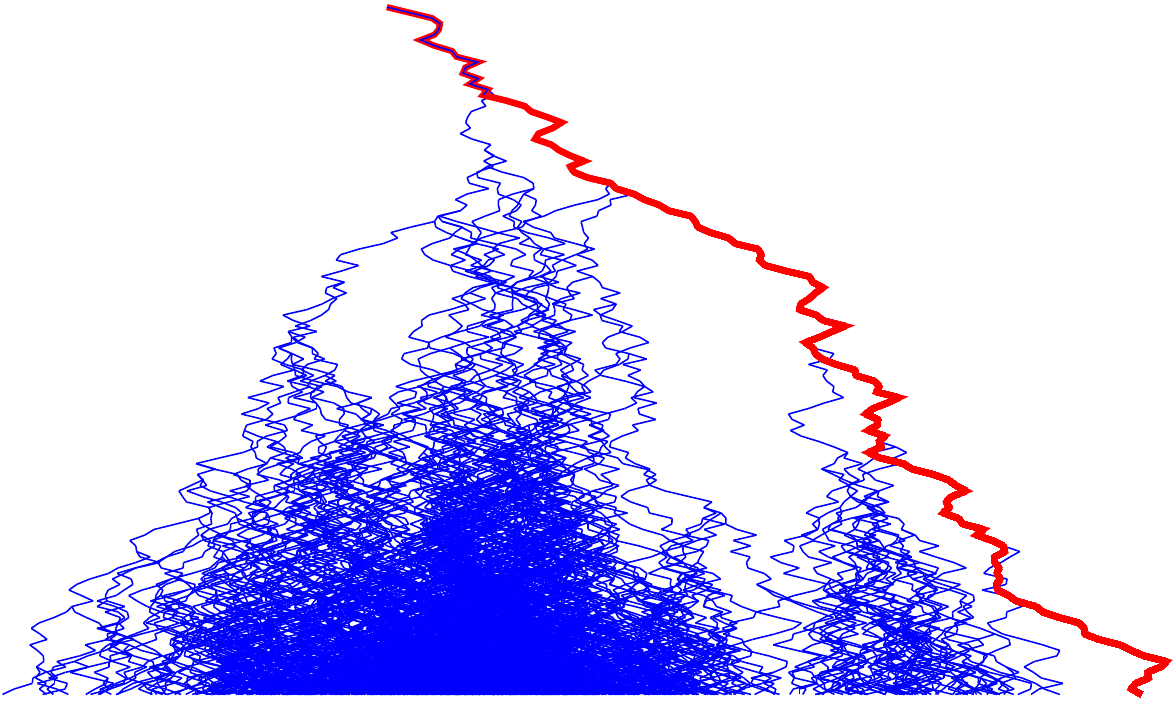}}
\caption{ A tree of the BBM which contributes to the large deviation function.  The  thick trajectory  is the trajectory of a red particle.
  This red particle is a particle which ends up on the right of position $v t$. This tree will contribute to the large deviation function of the $L$-BBM, $N$-BBM or CBRW if the red particle is not killed by one of the subtrees which branch from its trajectory.}
\end{figure}
\subsection{The $L$-BBM}
Consider first all the possible trees of a BBM which, starting with a single particle at the origin,  contain at least one particle which reaches, at time $t$, a position  on the right of $v t$ at time $t$.

Here we  focus on velocities $v > v_c$ (for the BBM one knows that $v_c = \sqrt{2} \sigma$).  The probability that the tree has at time $t$  at least one particle on the right of $v t$   is
  (\ref{ld-BBM},\ref{psi-BBM}) for $v > v_c$  
\begin{equation}
\label{pbbm}
P \sim \exp \left[t \left(1- {v^2 \over 2 \sigma^2} \right) \right]
= \exp \left[t \left(1- {v^2 \over v_c^2} \right) \right] \ .
\end{equation}
For each such  tree event,  we will call red particles all the particles  which end up on the right of $v t$.
Given its position at time $t$, the trajectory  of a   red particle is, up to a shift (linear in time), a Brownian bridge (in fact 
it is more like a Brownian excursion \cite{Chen,Aidekon} but this has no incidence on the discussion below). 

When one goes from the BBM to the $L$-BBM, a red particle will  survive  if between time $0$ and time $t$ no other particle of the BBM overtakes it by a distance $L$.
Any tree of the BBM for which a red particle survives   contributes to the event that the  the rightmost particle of $L$-BBM  is on the right of $v t$. So 
the probability that a tree of the BBM reaches position $vt$ and that at least one    red particle  is never overtaken by any other particle of the BBM by a distance $L$ is a lower bound for the probablity that a $L$-BBM reaches position $v t$. This is  why in the following, by estimating the  survival probability of  a  red particle  of a BBM, we will  get an upper  bound on the large deviation function (\ref{ld-LBBM})  of the $L$-BBM.

As a red particle is moving on average faster than $v_c$ the only possibility for it to be killed is that for a relatively short time  interval $s$, i.e. a time $s \ll t$, either this red particle moves slower than $v$, or one of the other particles of the tree moves sufficiently fast to overtake it by a distance $L$ or both.

So the picture is the following. A red particle moves at velocity $v$. Along its trajectory, branching events occur which give rise to subtrees.  This red particle is then killed if, shortly after   one of these branching events, the red particle slows down and one of the particles of the subtree overtakes it by a distance $L$. 

Let us now be quantitative.
The discussion below will hold for more general random walks, where the probability (\ref{pbbm})
would be replaced by
\begin{equation}
P \sim e^{t(1-f(v))}
\end{equation}
where 
  $f(v)$ is the large deviation function of the position of  the random walk.
In this general case $v_c$ is given by
\begin{equation}
f(v_c)=1 \ . 
\label{vc}
\end{equation}
 The case of the branching Brownian  motion will then be recovered by taking
\begin{equation}
f(v)= {v^2 \over 2 \sigma^2 }= {v^2 \over v_c^2} \ .
\label{f-BBM}
\end{equation}

One can  show that,  conditioned on the fact  that a red particle moves at  velocity $v$, the probability  $P(x,s)$ that  during a relatively short time interval  $(\tau, \tau+s)$ (here $1 \ll s \ll t$) it moves  a distance $x$  is
\begin{equation}
P(x,s) \sim \exp\left[ - s \left( f\left({x\over s}\right) - f(v) - \left({x\over s} - v\right) f'(v) \right) \right]  \ .
\label{Peq}
\end{equation}
Now the probability $Q(x,s)$ that at least one particle of the subtree created  at time $\tau$  moves a distance $x + L$ during the time interval $s$ is given by
\begin{equation}
Q(x,s)
\lesssim
 \min\left\{ 1, \exp\left[ s \left( 1 - f\left({x+L \over s}\right) \right) \right] \right\} \ .
\label{Qeq}
\end{equation}
Therefore the probability $p$ that such a  subtree will kill the red particle is
\begin{equation}
p \lesssim \max_{s,x} \{P(x,s) Q(x,s) \} \ .
\label{peq}
\end{equation}
{\it Remark:} It is rather easy to establish (\ref{Peq}). If a random walk has a large deviation function $f(v)$, the probability that during the time interval $(\tau, \tau+s)$ it moves  from a position $y$ to a  position $y+x$, conditioned on the fact that during on a time $t$ it moves a distance $v t$ (with $0 < \tau< \tau + s < t$) is given by
$${\exp[ -\tau f(y/\tau) - s f(x/s) -(t-\tau-s) f((vt-y-x)/(t-\tau-s))] \over \exp[t f(v)]}  \ . $$
Optimizing over $y$ gives
$$P(x,s)\sim {\exp[  - s f(x/s) -(t-s) f((vt-x)/(t-s))] \over \exp[t f(v)]} $$
and this leads to (\ref{Peq}) when $s \ll t$.

\ \\ 
Depending on   which term realizes the minimum in the rhs of (\ref{Qeq}) one
has to distinguish two cases:
\ \\ \ 
\begin{itemize}
\item If 1 dominates in (\ref{Qeq}) this means that
the particle of the subtree moves at velocity $v_c$.
In this case $x$ and $s$ are related by
\begin{equation}
 x+ L = v_c  \ s 
\label{front-position}
\end{equation}
because for $x < L- v_c s$, $Q(x,s)$ would remain $ \le 1$ but $P(x,s)$ would get smaller.
\\ \ \\ 
One can then see that the value of  $s$  which maximizes (\ref{peq}) is solution of
$$f\left(v_c - {L \over s} \right) - f(v) 
-\left(v_c - {L \over s} -v \right)  f'(v) 
 +{L\over s} f'\left(v_c - {L \over s} \right) 
 -{L\over s} f'\left(v \right) =0 \ . 
$$

This condition  takes the form 
\begin{equation}
f(y)-y f'(y) + v_c f'(y)=f(v) - v f'(v) + v_c f'(v)
\label{cond1}
\end{equation}
where $y=v_c-L/s$  and this gives  (\ref{peq})
\begin{equation}
\label{peq1}
p \sim e^{-L(f'(v)-f'(y))} \ . 
\end{equation}
Very much like  in the remark at the end of the introduction, assuming as above that $g(\lambda) \to \infty$ as $\lambda \to \pm \infty$, one can show that (\ref{cond1}) has always a solution. 
\\ \

As the number $B_t$ of branching events along the red trajectory is of order $t$ (for a rigorous justification, see Chauvin and Rouault~\cite{CR}) the survival probability of the red particle is
$$(1-p)^{B_t} \sim e^{-B_t p}$$
Therefore $$\P_{\rm LBBM}(X_{\rm max}(t) > v t) \gtrsim e^{t(1-f(v))-B_t p}$$
and this implies that
\begin{equation}
\label{peq2}
\psi_{\rm LBBM} - \psi_{\rm BBM} \lesssim p \sim e^{-L(f'(v)-f'(y))} \ . 
\end{equation}
In the particular case where $f(v)=v^2/(2\sigma^2)$   the solution of (\ref{cond1}) is  $y=2 v_c-v$   
 and this leads to the announced result (\ref{Res-LBBM},\ref{Res2-LBBM}).
\ \\ \ 
\item When the second alternative  dominates in (\ref{Qeq})  one needs to find the maximum over $s$ and $x$  of
$$
s \left[ -f\left({x\over s}\right) + f(v) + \left({x\over s} - v\right) f'(v) 
+ 1 - f\left({x+L \over s}\right)  \right] \ . 
$$

 This implies that $y=x/s$ and $s$ are solutions of
\begin{eqnarray}
&& f'(v) = f'(y) + f'\left( y + {L \over s} \right)   \nonumber \\
&& \label{cond2} \\
&& -f(y) + f(v) +(y-v) f'(v) +1 - f\left( y+{L\over s} \right) + {L \over s} f'\left(y + {L \over s} \right) =0 \ .
\nonumber
\end{eqnarray}
 After some algebra which uses (\ref{cond2}) one ends up with the same expression (\ref{peq2}), the only difference being that $y$ is now solution of (\ref{cond2}) instead of (\ref{cond1}).

As $v_c$ is solution of (\ref{vc}), one can check that the solution $y$ of (\ref{cond2})
reduces to the solution of (\ref{cond1}) when $y+ L/s \to v_c$,  meaning that the rightmost particle of the subtree moves at the velocity $v_c$.

In the particular case where $f(v)=v^2/(2 \sigma^2)$  the solution of (\ref{cond2}) is $y=(v - \sqrt{v^2-2 v_c^2})/2$  (where  $v_c=\sqrt{2} \sigma$) and this leads to the second line of (\ref{Res2-LBBM}).

\end{itemize}

\subsection{The $N$-BBM}
\label{subs:N-BBM}


In the $N$-BBM, the picture is rather similar and one has to estimate the probability $p$ that   a subtree will kill  a red particle. 
To do so  one needs the red particle  to slow down so that the subtree produces $N$ particles ahead of the red particle to eliminate it.

 The probability that the red particle moves a distance $x$ during time $s$ is still given by (\ref{Peq}).
We now need to estimate the probability  $Q(x,s)$ that the subtree 
produces, at  time $s$,  $N$ particles on  the right of postion $x$. 
We do not have an expression for $Q(x,s)$ (see the discusion
in the conclusion for a conjecture).
 One can however obtain
an easy upper bound  (using the Markov inequality)
$$Q(x,s) <  {\langle {\cal N}(x,s) \rangle \over N} $$
 where 
$ {\cal N}(x,s) $ is the number of particles of a subtree (of age $s$)  on the right of position $x$. One has  
$$ \langle {\cal N}(x,s) \rangle \sim \exp\left[s - {x^2 \over 2 \sigma^2 s} \right] $$ 
so that
\begin{equation}
Q(x,s) \lesssim \min\left[ 1, e^{s-\ln N  - {x^2 \over 2 \sigma^2 s}}\right]
\label{expression-justifiee}
\end{equation}
which, as for the $L$-BBM, we can  write  for more generality
\begin{equation}
Q(x,s) \lesssim \min\left[ 1, e^{s-\ln N  - s f({x\over s})}\right] 
\label{q-NBBM}
\end{equation}
to treat the case of an arbitrary $N$-BBM.

Now  we need to find a bound for $p$ given by (\ref{peq}) and the discussion is very similar to what we did for the $L$-BBM: 
\ \\ \ 
\begin{itemize}
\item If $1$ dominates in (\ref{q-NBBM}),  then $x=y s $ where $s$ and $y$ are related by 
\begin{equation}
 s-\ln N - s f (y) =0 \ .
\label{s-NBBM}
\end{equation}

The optimisation of (\ref{peq}) under the constraint (\ref{s-NBBM}) leads to 
$$p \sim e^{s[-f(y) + f(v) + (y -v) f'(v)]}$$
where    $y$ is solution of
\begin{equation}
 {1-f(y)+y f'(y)\over f'(y)}= 
 {1-f(v)+v f'(v)\over f'(v)} .
\label{Ber1}
\end{equation}

\noindent [A solution $y\not= v$ exists for $v>v_c$ for the same reason as in \eqref{v}.] One gets  after some algebra  
\begin{equation}
    p\sim N^{-\frac{f'(v)-f'(y)}{f'(y)}} \ .
\label{BBer1}
\end{equation}

For the $N$-BBM, one has $f(v)=v^2/(2 \sigma^2)$ the solution of (\ref{Ber1}) is  $y=2\sigma^2/v$; so $\frac{f'(v)-f'(y)}{f'(y)} = \frac{v^2}{2\sigma^2}-1$, and 
\begin{equation}
    p\sim N^{-(\frac{v^2}{2\sigma^2}-1)} \ .
    \label{Ber1bis}
\end{equation}

\noindent This agrees with the first line of (\ref{Res1-NBBM}).

\ \\ \ \\
\item In the second alternative of (\ref{q-NBBM})
\begin{equation}
p=\max_{s,y} \Big(\exp[s(1-f(y))-\ln  N +s(-f(y)+f(v)+(y-v)f'(v))]  \Big) \label{Ber2} \end{equation}
given that 
$s - \ln  N - s f(y) \le 0 \ . $
\noindent There is also the natural condition  $\ln N < s$ (because it is highly unlikely to have more than $e^s $ particles in a  time $(1-\varepsilon)s$, $\forall \varepsilon>0$) so that
\begin{equation}
s - s f(y) \le  \ln N  \le s . 
\label{Ber3}
\end{equation}

The expression in the exponential \eqref{Ber2} being linear in $s$, the maximum in $s$ is achieved at one of the two boundaries in \eqref{Ber3}. 

If the maximum is realized by the condition $s - s f(y) = \ln N$, one recovers the  results
 \eqref{BBer1}
 and \eqref{Ber1bis}. 
 On the other hand, if the maximum is realized by $s=\ln N$, the optimal value of $y$ in (\ref{Ber2}) is solution of 
\begin{equation}
 2f'(y)=f'(v)
\label{BBer3}
\end{equation}
and this leads to 
\begin{equation}
  p \sim N^{f(v)-vf'(v) - 2 f(y) + 2 y f'(y)}.
\label{BBer4}
\end{equation}
One can check that the range of validity of (\ref{BBer1}) is $v_c < v < v^*$ and for (\ref{BBer4})
is $v>v^*$ where $v^*$ is the value of $v$ where (\ref{Ber1}) and (\ref{BBer3}) have a common solution $y$. It is remarkable to notice that for $v=v^*$, both (\ref{BBer1}) and (\ref{BBer4}) coincide to give $p \sim N^{-1} $.

For $f(v)=v^2/(2 \sigma^2)$ the solution of  (\ref{BBer3}) is
 $y=v/2$, which leads to
\begin{equation}
    p \sim N^{- \frac{v^2}{4\sigma^2}} ;
    \label{Ber4}
\end{equation}

\noindent comparing \eqref{Ber1bis} with \eqref{Ber4}, one can check that \eqref{Ber1bis} holds for $v_c < v \le v^*=\sqrt{2} \, v_c = 2\sigma$, while \eqref{Ber4} is valid for $v \ge \sqrt{2} \, v_c$, as announced in (\ref{Res1-NBBM}).

\end{itemize}

\subsection{ The CBRW (branching random walk with coalescence)}
\label{subs:heuristic_CBRW}

For a branching random walk on a lattice, the probability that a red particle reaches the position $v t $  with $v > v_c$ at time $t$ is of the form
$$ e^{ t( r -       f(v))} \ . $$ 
For example if the random walk is characterized by the probability $\rho(y)$ that the walker jumps a distance $y$ from the site it occupies,  $f(v)$ is given in a parametric form as
\begin{equation}
f(v) = -g(\lambda) + \lambda g'(\lambda) \ \ \ \ \ ; \ \ \ \ v=g'(\lambda)
\label{parametric1}
\end{equation}
 with $g(\lambda)$ given by (\ref{gnew}).

Given that the red particle  moves on average at velocity $v$ during time $t$, the probability
$P(x,s)$  that it moves a distance $x$ during a time interval $1 \ll s \ll t$ is  as before (\ref{Peq}) by 
$$P(x,s) \sim  \exp \left[s \left(-f\left({x \over s }\right) + f(v) + \left({x \over s} - v \right) f'(v) \right)
\right] \ .  $$

On the other hand the number of particles produced by the subtree at position $x$ at time $s$ is $\lesssim e^{s (r - f(x/s))}$. Therefore  the probability that the red particle is killed by a subtree of age $s$ is
\begin{equation}
Q(x,s) 
\lesssim
\min \left[ 1,  \mu e^{s (r - f(x/s))}\right] \ .  
\label{Qexp}
\end{equation}

As for the $L$-BBM, one needs to distinguish two cases:
\ \\ \ 
\begin{itemize}
\item
If $1$ dominates in (\ref{Qexp}) this means that  $x/s$ satisfies
the relation 
\begin{equation}
 s \left( r -f\left(x \over s\right)\right) + \ln  \mu =0
\label{constraint}
\end{equation}
then one has to maximize  $P(x,s)$ given by (\ref{Peq}) over $s$ and $x$
given the constraint (\ref{constraint}).

This leads to  the  fact that $x= s y$ where $y$ is solution of 
\begin{equation}
{f(y)-r\over f'(y)} - y =
{f(v)-r\over f'(v)} - v 
\label{v1}
\end{equation}
and after some algebra to $p\sim \mu^{{f'(v)\over f'(y)}-1} $.
This leads to  (\ref{v}).
\ \\ \ 
\item
 The other case, when 
 (\ref{Qexp}) is  dominated by
$ \mu e^{s (r - f(x/s))} $, is much easier. The optimum over $s$ gives $s=0$ and therefore $p \sim \mu$.
\end{itemize}

\section{Proof: Existence of the large deviation function}
\label{existence}
In this section we prove the existence of the large deviation functions (\ref{ld-LBBM},\ref{ld-NBBM},\ref{ld-CBRW}).
\subsection{The $L$-BBM}
We first  establish two elementary properties of the $L$-BBM if one starts at time $0$ with $N$ surviving particles. In view of the statement, we can assume $N\ge 2$.

For any $s\ge 0$, let ${\cal N}(s)$ be the number of  surviving particles of the $L$-BBM at time $s$ (so  that  ${\cal N}(0)=N$).

\medskip

\begin{lemma}
\label{l:premiere_propriete_L-BBM}

 Let 
 \begin{equation}
\tau={a  \, L^2 \over 2  \sigma^2 \, \ln  N} \ .
\label{tau-def}
\end{equation}
Then
\begin{equation}
\P \left[ \exists s   \in (0,\tau]: {\cal N}(s)< N^\lambda \right] > 1 - { 3 \over N^\mu} \label{P1}
\end{equation}
where $a,b,\lambda$ and $\mu$ are constants which satisfy some conditions (\ref{a-b-lambda-nu}). For example, $a=36$, $b=3$, $\lambda= 17/18$ and $\mu=1/18$ will work. 
 
\end{lemma}

\begin{lemma}
\label{l:seconde_propriete_L-BBM}

 There exist constants $c_1>0$ and $c_2>0$, depending only on $(L, \, \sigma)$, such that 
 $$
 \P( \exists s\in [0, \, c_1] : \, {\cal N}(s) =1) 
 \ge
 c_2\ .
 $$
\end{lemma}

\medskip

In words, Lemma \ref{l:premiere_propriete_L-BBM} says that with a probability close to 1 when $N$ is large, the number of surviving particles ${\cal N}(\tau)$  will be greatly reduced within a very short time $\tau$ (defined in (\ref{tau-def})), whereas Lemma \ref{l:seconde_propriete_L-BBM} ensures that no matter how large $N$ is, within a time independent of $N$ (but which may depend on $L$ for example $L^2$), the total number of surviving particles will have become 1, at least once. In Lemma \ref{l:seconde_propriete_L-BBM}, it is possible to get moment estimates of the first time when the system has exactly a single particle; see \cite{Pain}.

\medskip

\noindent {\it Proof of Lemma \ref{l:premiere_propriete_L-BBM}.} It suffices to establish the following upper bound 
\begin{equation}
\P \left[ \forall s \in (0,\tau]: {\cal N}(s)\ge  N^\lambda \right]  <  { 3 \over N^\mu} \ .
\label{P1-bis}
\end{equation}
Let us write $$M=N^\lambda \ . $$ 
Without loss of generality, one can choose the origin to be the position of the rightmost particle of the $L$-BBM at time $0$. So all the initial positions are in $[-L, \ 0]$. 

 If we assume that   ${\cal N}(s)\ge M$ at all times $s < \tau$,  we want to follow  
the  trajectories $x_1(s) \cdots x_M(s)$   of $M$ surviving  particles between time $s=0$ and time $\tau$. At time $s=0$  we choose any set of    $M$  different particles  among  the $N$ present at time $0$. Let $x_1(0), \cdots x_M(0)$ be their positions at time $0$.
These particles move, branch and  can get  killed according to the rule of the $L$-BBM (they get killed  as soon as  their distance to the leading particle of the full $L$-BBM  exceeds $L$). When one of these $M$ particles gets killed, one replaces  it immediately by any of the remaining ${\cal N}(s)-(M-1)$.
On the other hand, when one of them branches, one just  keeps one of the two branches in our list of $M$ particles and ignore the other branch.
We obtain this way $M$ trajectories. 
Let us denote $x_1(s), \cdots x_M(s)$ the positions of these particles.
These $M$ trajectories  are those of Brownian particles, except that whever one of these particles gets killed, it is replaced by one of the surviving ${\cal N}(s)-M+1$ particles of the $L$-BBM (i.e. the corresponding trajectory makes a jump to its right).

Let us consider also $M$ regular Brownian motions which start at  time $s=0$ at the same positions as the above  $M$  particles of the $L$-BBM.  
We denote by $y_1(s) \cdots y_M(s)$ the positions of these $M$ Brownian particles at time $s$. By a simple coupling argument it is clear  that at any time $0<s<\tau$ and for $1 \le i \le M$, one has $y_i(s) \le x_i(s)$ so that   
$$\max_{1 \le i \le M} y_i(s)
 \le \max_{1 \le i \le M} x_i(s)  \ . $$
Therefore the probability  $Q$ that there exists  at least one surviving particle of the full $L$-BBM on the right of some fixed position $b L$ is bound from below
by
\begin{eqnarray*}
    Q 
 &\ge& \P \left[\max_{1 \le i \le M} x_i(s) > b L \right]
    \\
 &\ge& \P \left[\max_{1 \le i \le M} y_i(s) > b L \right]
    \ge
    1 - \left[\int_{-\infty}^{(b+1) L /\sqrt{2 \tau \sigma^2}} {e^{-u^2} du \over \sqrt{\pi} } \right]^M \ .
\end{eqnarray*}
Using the fact that for $x>2$
$$\int_{-\infty}^x {e^{-u^2} du \over \sqrt{\pi}} < 1 - e^{-2 x^2} < \exp[ - e^{- 2 x^2} ] $$
and that for $y >0$
$$ e^{-y} < {1 \over y}$$
one gets that
\begin{equation}
\label{Q1-bound}
Q > 1 - N^{ {2 (b+1)^2 \over a}- \lambda } \ .
\end{equation}
To complete the proof of (\ref{P1-bis}), we now show  that there is a small probability that  the number $\widehat{\cal N}$ of particles  of the $L$-BBM on the right of position $(b-1)L$  at time $\tau$  exceeds $M$.
To do so, we first notice  that
$$\P[ \widehat{\cal N} > M ] < 
\P[\widetilde{\cal N} > M ]  $$
where
$\widetilde{\cal N}    $ 
is  the number of particles on the right of $(b-1)L$  at time $\tau$ generated by    $N$ independent BBM's (with no selection)  starting  all  at time 0 at  position  $L$.
One can calculate the expectation 
$\widetilde{\cal N}    $ 
$$
\E[ \widetilde{\cal N}  ]    =   N e^\tau \int_{{(b-1) L \over \sqrt{2 \sigma^2 \tau}}}^\infty
{ e^{-u^2} du \over \sqrt{ \pi}}<
2 N^{1-{(b-1)^2 \over  a}}  $$
where we have used that for $x >0$
$$\int_x^\infty 
{ e^{-u^2} du \over \sqrt{ \pi}} < e^{-x^2} .$$

Therefore  by the Markov inequality one gets
$$\P[ \widehat{\cal N} > M ] < 
2  N^{1- \lambda -{(b-1)^2 \over  a}}  .$$

Now we know that, at time $\tau$,  there is a probability $Q$ close to $1$ that there is at least one particle on the right of $b L$ and  a probability also close to $1$ that   
$\widehat{\cal N} < M$.
Therefore, because when there is at least one particle on the left of $b L$ and no more than $M$ particles on the right of $(b-1) L$,  one knows that the total number of surviving particles of the $L$-BBM does not  exceed $M$. Consequently,
$$\P[{\cal N}(\tau) > M] < 1 - Q +  
\P[ \widehat{\cal N} > M ]   < 3
 N^{-\mu}$$
if we choose
\begin{equation}
\label{a-b-lambda-nu}
\mu= 
 -1+  \lambda +{(b-1)^2 \over  a} = 
\lambda - {2 (b+1)^2 \over a} .
\end{equation}
This completes the proof of  (\ref{P1}).\hfill$\Box$

\bigskip

\noindent {\it Proof of Lemma \ref{l:seconde_propriete_L-BBM}.} Let $C>0$ be a large constant independent of $N$. It suffices to prove that if one starts with  an arbitrary number $N$ of particles of the $L$-BBM, there is, uniformly in $N$,  a  positive probability $\widetilde{Q}$  that the number of particles will be less than or equal to $C$  at least once  before a time of order 1.

To prove this statement, we use $k = k(C, \, N)$ times the result (\ref{P1}): the number $k$ of steps needed is such that
$$
N^{\lambda^k} < C \le N^{\lambda^{k-1}} \ .
$$
%
%

\noindent According to (\ref{P1}),  one has 
 $$\widetilde{Q}> 
\left(1-{3 \over N^\mu} \right)
\cdots 
\left(1-{3 \over N^{\mu \lambda^{k-1}}} \right) 
\ge
\prod_{n=0}^\infty \Big( 1 - \frac{3}{C^{\mu \lambda^{-n}}} \Big) 
>0\ ,
$$

\noindent if the constant $C$ is chosen sufficiently large such that ${3 \over N^{\mu \lambda^{k-1}}} < 1$; on the other hand, the time needed (\ref{tau-def})  for this to happen  will be less than
$${  a L^2 \over 2 \sigma^2  } \sum_{n \ge 0} {\lambda^n \over \ln  C} 
={  a L^2 \over 2 \sigma^2  (1-\lambda) \ln  C} \ .$$
This proves Lemma \ref{l:seconde_propriete_L-BBM}.\hfill$\Box$

\bigskip

Now that we have proved Lemmas \ref{l:premiere_propriete_L-BBM} and \ref{l:seconde_propriete_L-BBM}, it is quite easy to deduce the existence of the large deviation function for the $L$-BBM. Let $v\in (-\infty, \, \infty)$, and let
$$
E_t
:=
\Big\{ \, \exists \hbox{ \rm particle in the $L$-BBM whose position at time $t$ is in } [vt, \, \infty) \Big\} \, .
$$

\noindent [Clearly, $E_t$ depends on $v$, $t$ and $N$.] The existence of the large deviation function we need to prove means the existence of $\lim_{t\to \infty} \frac1t \ln \P (E_t)$. We prove this by considering
$$
E_t^{(1)}
:=
\Big\{ {\cal N}(t) =1, \hbox{ \rm and the unique particle at time $t$ lies in } [vt, \, \infty) \Big\} \, ,
$$

\noindent where ${\cal N}(t)$ denotes as before the number of particles in the $L$-BBM at time $t$. Clearly,
$$
\P (E_{t+t'}^{(1)}) 
\ge 
\P (E_t^{(1)})\, \P (E_{t'}^{(1)}) \, , 
\qquad
\forall t\ge 0, \; t'\ge 0\, .
$$

\noindent As such, the function $t\mapsto \ln \P (E_t^{(1)})$ is superadditive on $(0, \, \infty)$, and as $t$ goes to infinity, $\frac1t \ln \P (E_t^{(1)}) \to \sup_{s>0} \frac1s \ln \P (E_s^{(1)})\in (-\infty, \, 0]$.

The existence of $\lim_{t\to \infty} \frac1t \ln \P (E_t^{(1)})$ implies the existence of $\lim_{t\to \infty} \frac1t \ln \P (E_t)$; indeed, we trivially have
$$
\P (E_t)
\ge
\P (E_t^{(1)}) ,
\qquad \forall t>0 \, ,
$$

\noindent because $E_t \supset E_t^{(1)}$. Conversely, by Lemma\ref{l:seconde_propriete_L-BBM}, 
$$
\P (E_{t+c_1}^{(1)})
\ge
c_2\, \P (E_t) ,
\qquad \forall t>0 \ .
$$

\noindent The last two inequalities together yield the existence of $\lim_{t\to \infty} \frac1t \ln \P (E_t)$, which equals $\lim_{t\to \infty} \frac1t \ln \P (E_t^{(1)})$.

\subsection{The $N$-BBM}

For the $N$-BBM, we start with two simple but useful monotonicity properties, which are borrowed from \cite{DR}. We include the elementary proof for the sake of self-containedness. We say that $(u_i)_{1\le i\le M}$ dominates $(v_i)_{1\le i\le N}$ if $\sum_{i=1}^M {\bf 1}_{\{ u_i \ge a\} } \ge \sum_{i=1}^N {\bf 1}_{\{ v_i \ge a\} }$ for all $a\in (-\infty, \, \infty)$ (so in particular, $M\ge N$).

\medskip

\begin{lemma}
\label{l:premiere_propriete_de_monotonie_N-BBM}

 {\bf (First monotonicity property for the $N$-BBM)}
 Let $x_1 \ge \cdots \ge x_N$ and $y_1 \ge \cdots \ge y_N$ be such that $x_i \ge y_i$ for all $1\le i\le N$. There exists a coupling for two $N$-BBM systems on a same probability space, starting at positions $(x_i)_{1\le i\le N}$ and $(y_i)_{1\le i\le N}$ respectively, such that the first system dominates the second at all time.
\end{lemma}

\medskip

\noindent {\it Proof.} Consider two $N$-BBM systems, the first starting at positions $(x_i)_{1\le i\le N}$, and the second at $(y_i)_{1\le i\le N}$. We attach the same Brownian motion to particles starting at $x_i$ and $y_i$ (for $1\le i\le N$) respectively in the two systems, and also attach the same Poisson process which determines the branching times along the paths. As such, the first branching time is identical in the two systems, and before this time, the $x$-system obviously dominates the $y$-system. It is also easy to check that right after the first branching time, the $x$-system still dominates the $y$-system. Then by attaching as before the same Brownian motions and the same Poissonian clocks to the $x$- and the $y$-particles, the $x$-system will continue to dominate the $y$-system. And so on. The procedure leads to the desired coupling.\hfill$\Box$

\medskip

\begin{lemma}
\label{l:seconde_propriete_de_monotonie_N-BBM}

 {\bf (Second monotonicity property for the $N$-BBM)}
 Let $N'\ge N$. Let $x_1 \ge \cdots \ge x_{N'}$ and $y_1 \ge \cdots \ge y_N$ be such that $x_i \ge y_i$ for all $1\le i\le N$. There exists a coupling for an $N'$-BBM and an $N$-BBM on a same probability space, with initial positions $(x_i)_{1\le i\le N'}$ and $(y_i)_{1\le i\le N}$ respectively, such that the $N'$-BBM dominates the $N$-BBM all time.

\end{lemma}

\medskip

\noindent {\it Proof.} If $N'=N$, this amounts to the previous lemma. So let us assume $N'>N$. Then, as in the proof of the previous lemma, if initially the $N$ rightmost particles of the system with $N'$ particles dominates the other system, this remains true subsequently. The remaining $N'-N$ particles can only reinforce this domination.\hfill$\Box$

\bigskip

Let us now turn to the proof of the existence of the large deviation function for the $N$-BBM. Let $v\in R$. Consider the following event for the $N$-BBM:
$$
E_t
:=
\Big\{ \, \exists \hbox{ \rm particle whose position at time $t$ lies in } [vt, \, \infty) \Big\} \, .
$$

\noindent To prove the existence of the large deviation function, we need to show that the limit $\lim_{t\to \infty} \frac1t \ln \P (E_t)$ exists. We prove this by an argument of superadditivity. By removing all particles at time $t$ except the rightmost one, the second monotonicity property stated in Lemma \ref{l:seconde_propriete_de_monotonie_N-BBM} tells us that  
$$
\P (E_{t+t'})
\ge
\P (E_t) \, \P (E_{t'}),
\qquad \forall t>0, \, \forall t'>0 \, .
$$

\noindent So the function $t\mapsto \ln \P (E_t)$ is superadditive on $(0, \, \infty)$. In particular,
$$
\lim_{t\to \infty} \, \frac1t \ln \P (E_t)
=
\sup_{t>0} \, \frac1t \ln \P (E_t)
\in
(-\infty, \, 0] \, ,
$$

\noindent exists. 

\subsection{ The CBRW (branching random walk with coalescence)}

The existence of the large deviation function of the CBRW is very similar. As in Lemma \ref{l:seconde_propriete_de_monotonie_N-BBM} for the $N$-BBM, the probability of the large deviation event increases with the number of initial particles. Consequently, by removing all particles except the rightmost one at time $t$, on sees that if $E_t$ denotes the event that in the CBRW, there exists a particle lying in $[vt, \  \infty)$ at time $t$, 
$$
\P (E_{t+t'})
\ge
\P (E_t) \, \P (E_{t'}),
\qquad \forall t>0, \, \forall t'>0 \, ,
$$

\noindent from which the existence of $\lim_{t\to \infty} \frac1t \ln \P (E_t)$ follows immediately.

\section{Proof: Bounds for the large deviation function}
\label{proofs}

\subsection{General strategy}

We first describe the strategy for the $L$-BBM. The strategy for the $N$-BBM will be along similar lines, with a few appropriate modifications indicated below. The output of this paragraph has been described in Section \ref{physical}.

Let $E_t^{\mathrm{LBBM}}$ denote as before the event that there exists at least one particle in the $L$-BBM whose position at time $t$ lies in $[vt, \, \infty)$. To bound from below $\P (E_t^{\mathrm{LBBM}})$, we consider the following event of the BBM (without selection):\footnote{Although the right-hand side of \eqref{event_L} is an event of the BBM, not of the $L$-BBM, we use the superscript in $\widetilde{E}_t^{\mathrm{LBBM}}$ to remind us that it will serve to study the large deviation function for the $L$-BBM. A similar remark applies to the forthcoming events $\widetilde{E}_t^{\mathrm{NBBM}}$ and $\widetilde{E}_t^{\mathrm{CBRW}}$.}
\begin{eqnarray}
    \widetilde{E}_t^{\mathrm{LBBM}}
 &:=&
    \bigcup_{i=1}^{{\cal N}(t)}\, \{ \hbox{\rm the particle $i$ lies in $[vt, \, \infty)$ at time $t$,} 
    \nonumber
    \\
 &&\qquad\qquad \hbox{\rm not $L$-dominated, and leans to the left} \} \, .
    \label{event_L}
\end{eqnarray}

\noindent Here, ${\cal N}(t)$ denotes, as before, the number of particles at time $t$. Leaning to the left means that the path of the particle lies in $(-\infty, \, t'v+ t^{2/3}]$ for all $t'\in [0, \, t]$.\footnote{The choice of the power $2/3$ is arbitrary; anything in $(\frac12, \, 1)$ will do the job.} 
We say that a particle with trajectory $(X_{t'}, \, t'\in [0, \, t])$ is $L$-dominated if at some time $t' \in [0, \, t]$ there is a particle lying in $[X_{t'} + L, \, \infty)$. 

Clearly, if $\widetilde{E}_t^{\mathrm{LBBM}}$ is realized, then one can construct an $L$-BBM such that the large deviation event $E_t^{\mathrm{LBBM}}$ is realized. Therefore,
$$
\P (\widetilde{E}_t^{\mathrm{LBBM}}) \le \P (E_t^{\mathrm{LBBM}})\, .
$$

We estimate $\P (\widetilde{E}_t^{\mathrm{LBBM}})$ which will serve as a lower bound for $\P (E_t^{\mathrm{LBBM}})$. To bound $\P (\widetilde{E}_t^{\mathrm{LBBM}})$ from below, let us write
\begin{equation}
    \# \widetilde{E}_t^{\mathrm{LBBM}}
    :=
    \sum_{i=1}^{{\cal N}(t)}\, {\bf 1}_{ \{ \hbox{\scriptsize\rm the particle $i$ lies in $[vt, \, \infty)$ at time $t$, not $L$-dominated, and leans to the left} \} } \, .
    \label{|event_L|}
\end{equation}

\noindent By the Cauchy--Schwarz inequality, we have
$$
\P (\widetilde{E}_t^{\mathrm{LBBM}}) 
\ge 
\frac{[\E (\# \widetilde{E}_t^{\mathrm{LBBM}})]^2}{\E [(\# \widetilde{E}_t^{\mathrm{LBBM}})^2]} \, .
$$

\noindent A fortiori,
\begin{equation}
    \P (E_t^{\mathrm{LBBM}})
    \ge
    \frac{[\E (\# \widetilde{E}_t^{\mathrm{LBBM}})]^2}{\E [(\# \widetilde{E}_t^{\mathrm{LBBM}})^2]} \, .
    \label{cauchy-schwarz}
\end{equation}

\noindent We need to bound $\E(\# \widetilde{E}_t^{\mathrm{LBBM}})$ from below, and bound $\E[(\# \widetilde{E}_t^{\mathrm{LBBM}})^2]$ from above. The main estimates for the $L$-BBM which we obtain below are as follows:
\begin{eqnarray}
    \E (\# \widetilde{E}_t^{\mathrm{LBBM}})
 &\gtrsim& \exp \left[ - \Big( \frac{v^2}{2\sigma^2} -1 + e^{-[\alpha(v)+o_L(1)] \, L} \Big) t \right] \, ,
    \label{L-BBM:E(Et):lb}
    \\
    \E [(\# \widetilde{E}_t^{\mathrm{LBBM}})^2]
 &\lesssim& \exp \left[ -\Big( \frac{v^2}{2\sigma^2} -1\Big) t \right] ,
    \label{L-BBM:E(Et):ub}
\end{eqnarray}

\noindent with the value of $\alpha(v)$ given in \eqref{Res2-LBBM}. As before, the notation $a(t) \gtrsim b(t)$ or $b(t) \lesssim a(t)$ means that $\liminf_{t\to \infty} \frac1t \ln (\frac{a(t)}{b(t)}) \ge 0$, whereas $o_L(1)$ denotes a term not depending on $t$, such that $\lim_{L\to \infty} o_L(1) =0$. In view of the Cauchy--Schwarz inequality \eqref{cauchy-schwarz}, it is clear that \eqref{L-BBM:E(Et):lb} and \eqref{L-BBM:E(Et):ub} together will imply the upper bound stated in \eqref{Res-LBBM} for the large deviation function $\psi_{\mathrm{LBBM}}$ of the $L$-BBM.

The next subsection is devoted to the proof of (\ref{L-BBM:E(Et):lb}). The proof of (\ref{L-BBM:E(Et):ub}), which is identical for all the three models, is postponed to Subsection \ref{2nd_moment}.

\subsection{First moment computations for the $L$-BBM}

We write $X=(X_u, \, u\in [0, \, t])$ for the trajectory of the particle $i$ in the definition of $\# \widetilde{E}_t^{\mathrm{LBBM}}$, and write 
\begin{equation}
    A_t
    :=
    \{ X_u \le uv+t^{2/3}, \, \forall u\in [0, \, t]\} \ ,
    \label{At}
\end{equation}

\noindent which stands for the event that the particle $i$ leans to the left. Then
\begin{equation}
    \E (\# \widetilde{E}_t^{\mathrm{LBBM}})
=
\int_{tv}^\infty \frac{e^{t-\frac{y^2}{2\sigma^2 t}}}{(2\pi \sigma^2 t)^{1/2}} \, 
\E \Big( {\bf 1}_{A_t}\prod_{j: \, \tau_j \le t} {\bf 1}_{D_t^{\mathrm{LBBM}}(\tau_j)} \, \Big| \, X_t = y\Big) \, \mathrm{d} y \, ,
    \label{premier-moment}
\end{equation}

\noindent where, for all $u\in [0, \, t]$, $D_t^{\mathrm{LBBM}}(u)$ stands for the event that the subtree of BBM branched at time $u$ on the path of $X$ does not produce any descendant going beyond $X$ by distance $\ge L$ at any time during $[u, \, t]$. Here, $(\tau_j, \, j\ge 1)$ is a rate-2 Poisson process. The identity above, which is intuitively clear (except, maybe, for the rate being 2 instead of 1 which is a property of the Poisson process; we mention that the rate of the Poisson process plays no role in the final result), follows immediately from the Chauvin--Rouault spinal decomposition theorem \cite{CR}.

It is easily guessed
that the essential contribution to the integral $\int_{tv}^\infty \cdots \, \mathrm{d} y$ on the right-hand side comes from the neighbourhood of $y=vt$. In any case, we can limit ourselves to the neighbourhood of $y=vt$ to pretend that it only gives a lower bound: 
$$
\E (\# \widetilde{E}_t^{\mathrm{LBBM}})
\gtrsim
e^{-(\frac{v^2}{2\sigma^2} -1)t}\, 
\E \Big( {\bf 1}_{A_t} \prod_{j: \, \tau_j \le t} {\bf 1}_{D_t^{\mathrm{LBBM}}(\tau_j)} \, \Big| \, X_t = vt\Big) \, .
$$

\noindent By conditioning upon $X:= (X_u, \, u \in [0, \, t])$ and $\tau := (\tau_j, \, j\ge 1)$, we have
$$
\E \Big( {\bf 1}_{A_t} \prod_{j: \, \tau_j \le t} {\bf 1}_{D_t^{\mathrm{LBBM}}(\tau_j)} \, \Big| \, X, \, \tau\Big) 
=
{\bf 1}_{A_t} \prod_{j: \, \tau_j \le t} \P_X (D_t^{\mathrm{LBBM}}(\tau_j)\, | \, \tau)\, ,
$$

\noindent where $\P_X (\, \cdot \, ) := \P (\, \cdot \, | \, X)$ denotes conditional probability given $X$. As such, writing $\E_X$ for expectation with respect to $\P_X$, we have
\begin{eqnarray*}
    \E_X \Big( {\bf 1}_{A_t} \prod_{j: \, \tau_j \le t} {\bf 1}_{D_t^{\mathrm{LBBM}}(\tau_j)}\Big)
 &=& {\bf 1}_{A_t} \, \E_X \Big( \prod_{j: \, \tau_j \le t} \P_X (D_t^{\mathrm{LBBM}}(\tau_j)\, | \, \tau) \Big)
    \\
 &=& {\bf 1}_{A_t} \, e^{- 2\int_0^t [1-\P_X (D_t^{\mathrm{LBBM}}(u))] \, \mathrm{d} u} \, ,
\end{eqnarray*}

\noindent the second identity being a consequence of the fact that $(\tau_i, \, i\ge 1)$ is a rate-2 Poisson process. Accordingly,
$$
\E (\# \widetilde{E}_t^{\mathrm{LBBM}})
\gtrsim
e^{-(\frac{v^2}{2\sigma^2} -1)t}\, 
\E \Big\{ {\bf 1}_{A_t} \, e^{- 2\int_0^t [1-\P_X (D_t^{\mathrm{LBBM}}(u))] \, \mathrm{d} u} \, \Big| \, X_t = vt\Big\} \, .
$$

\noindent Given $X_t = vt$, the process $(X_u, \, u\in [0, \, t])$ is a Brownian bridge of length $t$; it can be realized as $X_u = vu + \sigma(W_u - \frac{u}{t}W_t)$, where $W$ is a standard Brownian motion (of variance $1$). Thus
\begin{equation}
    \E (\# \widetilde{E}_t^{\mathrm{LBBM}})
    \gtrsim
    e^{-(\frac{v^2}{2\sigma^2} -1)t}\, 
    \E \Big\{ {\bf 1}_{A_t^{(W)}} \, 
e^{- 2\int_0^t [1-\P_X (D_t^{\mathrm{LBBM}}(u))] \, \mathrm{d} u} \Big\} .
    \label{pf-lb-1}
\end{equation}

\noindent where
$$
A_t^{(W)}
:=
\{ W_u - \frac{u}{t}W_t \le \frac{t^{2/3}}{\sigma} , \, \forall u\in [0, \, t]\} .
$$

We will see that the indicator ${\bf 1}_{A_t^{(W)}}$ brings no significant difference to the expectation. Writing the conditional probability
$$
\P^t (\, \cdot \, )
:=
\P\Big( \, \cdot \, \Big| \, A_t^{(W)} \Big) ,
$$

\noindent and $\E^t (\, \cdot \, )$ for the associated expectation, we obtain:
$$
\E (\# \widetilde{E}_t^{\mathrm{LBBM}}) 
\gtrsim
e^{-(\frac{v^2}{2\sigma^2} -1)t}\, \P (A_t^{(W)}) \,
\E^t \Big( e^{- 2\int_0^t [1-\P_X (D_t^{\mathrm{LBBM}}(u))] \, \mathrm{d} u} \Big) \, .
$$

\noindent By scaling, $\P(A_t^{(W)}) = \P\{ W_r - r\, W_1 \le \frac{t^{1/6}}{\sigma} , \, \forall r\in [0, \, 1]\}$, which converges to $1$ when $t\to \infty$. So in our notation for ``$\gtrsim$", we have
\begin{eqnarray*}
    \E (\# \widetilde{E}_t^{\mathrm{LBBM}}) 
 &\gtrsim& e^{-(\frac{v^2}{2\sigma^2} -1)t}\, \E^t \Big( e^{- 2\int_0^t [1-\P_X (D_t^{\mathrm{LBBM}}(u))] \, \mathrm{d} u} \Big)
    \\
 &\ge& \exp \Big\{ -(\frac{v^2}{2\sigma^2} -1)t - 2\int_0^t \E^t [1-\P_X (D_t^{\mathrm{LBBM}}(u))] \, \mathrm{d} u \Big\}\, ,
\end{eqnarray*}

\noindent the last line following from Jensen's inequality. By definition,
\begin{eqnarray*}
    \E^t [1-\P_X (D_t^{\mathrm{LBBM}}(u))]
 &=& \frac{\E \{ [1-\P_X (D_t^{\mathrm{LBBM}}(u))] \, {\bf 1}_{A_t^{(W)}}\}}{\P(A_t^{(W)})}
    \\
 &\le& \frac{\E [1-\P_X (D_t^{\mathrm{LBBM}}(u))]}{\P(A_t^{(W)})} .
\end{eqnarray*}

\noindent We have already seen that $\P(A_t^{(W)}) \to 1$, $t\to \infty$. So for all sufficiently large $t$ (which will be taken for granted from now on), we have\footnote{The choice of $3$ on the right-hand side is arbitrary; anything in $(2, \, \infty)$ will do the job.} 
$$
\frac{2}{\P(A_t^{(W)})} \le 3.
$$

\noindent As such,
\begin{equation}
    \E (\# \widetilde{E}_t^{\mathrm{LBBM}}) 
    \gtrsim
    \exp \Big\{ -(\frac{v^2}{2\sigma^2} -1)t - 3\int_0^t \E [1-\P_X (D_t^{\mathrm{LBBM}}(u))] \, \mathrm{d} u \Big\}\, .
    \label{L-BBM:Jensen}
\end{equation}

\noindent [So the presence of the indicator function ${\bf 1}_{A_t^{(W)}}$ in \eqref{pf-lb-1} indeed has no significant influence.]

For all $s>0$, let us write $M(s)$ for the maximal position at time $s$ of a BBM independent of $X$. [This was denoted by $X_{\max}(s)$ in the introduction.] By definition of $D_t^{\mathrm{LBBM}}(u)$,
\begin{eqnarray}
    1-\P_X (D_t^{\mathrm{LBBM}}(u))
 &=&\P_X \Big( \exists s\in (0, \, t-u]: \, M(s) \ge L+X_{s+u}-X_u \Big)
    \nonumber
    \\
 &\le& \int_0^{t-u} \P_X \Big( M(s) \ge L+X_{s+u}-X_u \Big) \, \mathrm{d} s \, .
    \label{premier_moment_L-BBM:estimation_technique}
\end{eqnarray}

\noindent [The inequality in \eqref{premier_moment_L-BBM:estimation_technique} is heuristic; it would be trivially true if $s$ were an integer (in which case we would have a sum over $s$ instead of an integral on the right-hand side). 
However, we can easily make it rigorous by arguing that $\P_X ( \exists s\in (0, \, t-u]: \, M(s) \ge L+X_{s+u}-X_u ) \le \sum_{i=1}^{\lfloor t-u\rfloor+1} \P_X ( \sup_{s\in [i-1, \, i]} M(s) \ge L+ \inf_{s\in [i-1, \, i]} (X_{s+u}-X_u ))$. The rest of the argument will go through, by noting that the tail probability of $\sup_{s\in [i-1, \, i]} M(s)$ behaves like the tail probability of $M(i)$ (in the sens of "$\lesssim$"), and that in the estimates of $J_t^{(1)}(u, \, s)$ and $J_t^{(1)}(u, \, s)$, instead of using the exact Gaussian distribution of $W_{s+u}-W_u -\frac{s}{t} W_t$, we can use the fact that the negative tail distribution of $\inf_{s\in [i-1, \, i]}	(W_{s+u}-W_u -\frac{s}{t} W_t)$ is bounded by the Gaussian tail.
The same argument applies to the $N$-BBM. For the CBRW, the situation is slightly different due to the fact that the space is discrete, but some obvious modifications to the argument readily make it rigorous.]
 
By the Markov inequality, $\P_X \{M(s) \ge L+X_{s+u}-X_u\}$ is bounded by the $\P_X$-expectation of the number of particles located beyond $L+X_{s+u}-X_u$ at time $s$; this $\P_X$-expectation is 
bounded by $\exp ( s - \frac{(L+X_{s+u}-X_u)^2}{2\sigma^2 s})$. Of course, this bound is interesting only when $L+X_{s+u}-X_u \ge (2\sigma^2)^{1/2} s$; otherwise, we use the trivial inequality $\P_X \{ M(s) \ge L+X_{s+u}-X_u\} \le 1$. As a consequence,
\begin{eqnarray*}
    1-\P_X (D_t^{\mathrm{LBBM}}(u))
 &\le& \int_0^{t-u} \Big[ {\bf 1}_{\{ L+X_{s+u}-X_u < (2\sigma^2)^{1/2} s\} } + 
    \\
 &&\hskip-20pt +{\bf 1}_{\{ L+X_{s+u}-X_u \ge (2\sigma^2)^{1/2} s\} } \exp\Big( s - \frac{(L+X_{s+u}-X_u)^2}{2\sigma^2 s} \Big) \Big] \, \mathrm{d} u \, .
\end{eqnarray*}

\noindent With the notation $X_u = vu + \sigma(W_u - \frac{u}{t}W_t)$, we have $L+X_{s+u}-X_u = L+ vs + \sigma (W_{s+u}-W_u -\frac{s}{t} W_t)$. Assembling these pieces yields that
$$
\E (\# \widetilde{E}_t^{\mathrm{LBBM}})
\gtrsim
\exp\Big\{ -(\frac{v^2}{2\sigma^2} -1)t -3 \int_0^t  \Big( \int_0^{t-u} [J_t^{(1)}(u, \, s) + J_t^{(2)}(u, \, s)] \, \mathrm{d} s \Big) \, \mathrm{d} u \Big\} \, ,
$$

\noindent where
\begin{eqnarray*}
    J_t^{(1)}(u, \, s)
 &:=&\P \Big( L+ vs + \sigma (W_{s+u}-W_u -\frac{s}{t} W_t) < (2\sigma^2)^{1/2} s \Big) \, ,
    \\
    J_t^{(2)}(u, \, s)
 &:=&\E \Big[ {\bf 1}_{\{ L+ vs + \sigma (W_{s+u}-W_u -\frac{s}{t} W_t) \ge (2\sigma^2)^{1/2} s\} } \times
    \\
 && \times \exp\Big( s - \frac{[L+ vs + \sigma (W_{s+u}-W_u -\frac{s}{t} W_t)]^2}{2\sigma^2 s} \Big) \Big].
\end{eqnarray*}

\noindent The random variable $W_{s+u}-W_u -\frac{s}{t} W_t$ has the Gaussian ${\cal N}(0, \, s(1-\frac{s}{t}))$ law. Some elementary but tedious computations lead to the following conclusion: in case $v>(\frac{9\sigma^2}{2})^{1/2}$, the subtrees move forward faster than the usual speed $(2\sigma^2)^{1/2}$ (i.e., the integral of $J_t^{(2)}(u, \, s)$ dominates), whereas if $(2\sigma^2)^{1/2} <v \le (\frac{9\sigma^2}{2})^{1/2}$, these subtrees make no particular effort: they only need, in this case, to wait for the occasions when the red particle makes some fluctuations toward the left (which happens with some frequency). Letting $t\to \infty$ and then $L\to \infty$ (in this order), we obtain:
$$
    \E (\# \widetilde{E}_t^{\mathrm{LBBM}})
    \ge
    \exp\left[ - (1+o(1)) (\frac{v^2}{2\sigma^2} - 1 + e^{- (1+o_L(1)) \alpha(v) L}) \, t\right] \, ,
    \label{premier-moment:lb}
$$

\noindent where $\alpha(v)$ is given in \eqref{Res2-LBBM}. This is the desired lower bound \eqref{L-BBM:E(Et):lb}.

\subsection{First moment computations for the $N$-BBM}

The proof for the $N$-BBM is similar to the proof for the $L$-BBM, so we present only an outline, indicating the places where modifications are needed. We fix $0<\varepsilon<1$, and write $M = M(\varepsilon) := \lfloor N^{1-\varepsilon} \rfloor$. Consider
\begin{eqnarray*}
    \widetilde{E}_t^{\mathrm{NBBM}}
 &:=&\bigcup_{i=1}^{{\cal N}(t)}\, \{ \hbox{\rm particle $i$ lies in $[vt, \, \infty)$, leans to the left,} 
    \\
 &&\qquad\qquad \hbox{\rm does not split much, is not $M$-dominated} \} \, .
\end{eqnarray*}

\noindent Let us explain the definition of $\widetilde{E}_t^{\mathrm{NBBM}}$. The meaning of "leans to the left" is as for the $L$-BBM: the path of the particle lies in $(-\infty, \, t'v+t^{2/3}]$ for all $t'\in [0, \, t]$. By "does not split much", we mean\footnote{The choice of powers in $(\ln N)^2$ and $(\ln N)^3$ are arbitrary: they can be replaced by $C_1 \ln N$ and $C_2 \ln N$ with two sufficiently large constants $C_1$ and $C_2$.} that the number of branchings (from the path of the particle $i$) at each of the time intervals $[(k-1)(\ln N)^2, \, k (\ln N)^2]$, for $1\le k\le \frac{t}{(\ln N)^2}$, is bounded by $(\ln N)^3$. By "$M$-dominated", we mean the existence of a time $u\in [0, \, t]$ such that either there are at least $M$ particles branching at time $u$ from the path of the particle $i$ lying in $[X_{t'}, \, \infty)$ at some time $t' \in [u, \, u+(\ln N)^2]$, or there is a particle branching at time $u$ from the path of the particle $i$ lying in $[X_{t'}, \, \infty)$ at some time $t' \in [u+(\ln N)^2, \, t]$ (if the interval is not empty).

The event $\widetilde{E}_t^{\mathrm{NBBM}}$ is the analogue, for the $N$-BBM, of the event $\widetilde{E}_t^{\mathrm{LBBM}}$ in \eqref{event_L}. The probability $\P(\widetilde{E}_t^{\mathrm{NBBM}})$ will serve as a lower bound for the probability of the large deviation event for the $N$-BBM, because by definition, $\widetilde{E}_t^{\mathrm{NBBM}}$ implies the large deviation event for the $N$-BBM.

Write as before $\# \widetilde{E}_t^{\mathrm{NBBM}}$ for the number of $i$ satisfying the conditions in $\widetilde{E}_t^{\mathrm{NBBM}}$. The main estimates for the $N$-BBM we are going to prove are: 
\begin{eqnarray}
    \E(\# \widetilde{E}_t^{\mathrm{NBBM}})
 &\gtrsim& \exp\left[ -\Big( \frac{v^2}{2\sigma^2} -1 + M^{-\beta(v)+o_N(1)} \Big) t \right] \, ,
    \label{N-BBM:E(Et):lb}
    \\
    \E[(\# \widetilde{E}_t^{\mathrm{NBBM}})^2]
 &\lesssim& \exp\left[ - \Big(\frac{v^2}{2\sigma^2} -1 \Big)t\right] \, ,
    \label{N-BBM:E(Et):ub}
\end{eqnarray}

\noindent where $\beta(v)$ is defined in \eqref{Res1-NBBM}, and $o_N(1)$ stands for a term not depending on $t$ such that $\lim_{N\to \infty} o_N(1) =0$. Since $\varepsilon$ can be as small as possible, \eqref{N-BBM:E(Et):lb} and \eqref{N-BBM:E(Et):ub} together with the Cauchy--Schwarz inequality will yield the upper bound stated in \eqref{Res-NBBM} for the large deviation function for the $N$-BBM.

The proof of (\ref{N-BBM:E(Et):ub}), which is identical for all the three models, is postponed to Subsection \ref{2nd_moment}. The rest of this subsection is devoted to the proof of \eqref{N-BBM:E(Et):lb}. 

Writing $X=(X_u, \, u\in [0, \, t])$ again for the trajectory of the red particle $i$, and $A_t := \{ X_u \le uv+t^{2/3}, \, \forall u\in [0, \, t]\}$ as in \eqref{At}, we have
$$
\E(\# \widetilde{E}_t^{\mathrm{NBBM}})
=
\int_{tv}^\infty \frac{e^{t-\frac{y^2}{2\sigma^2 t}}}{(2\pi \sigma^2 t)^{1/2}} \, 
\E \Big( {\bf 1}_{A_t} \, (\prod_{k=1}^{t/(\ln N)^2} {\bf 1}_{G_k})
\prod_{j: \, \tau_j \le t} {\bf 1}_{D_t^{\mathrm{NBBM}}(\tau_j)} \, \Big| \, X_t = y\Big) \, \mathrm{d} y \, ,
$$

\noindent where, for all $u\in [0, \, t]$, $D_t^{\mathrm{NBBM}}(u)$ stands for the event that the subtree of BBM branched at time $u$ on the path of $X$ does not produce $M$ descendants going beyond $X$ at any time during $[u, \, u + (\ln N)^2]$ and does not produce any descendant going beyond $X$ at any time during $[u + (\ln N)^2, \, t]$ (if the interval is non empty). Here, $(\tau_j, \, j\ge 1)$ is as before the atoms of a rate-2 Poisson process, and for each $k$, $G_k$ is the event that the number of atoms $(\tau_j, \, j\ge 1)$ lying in $[(k-1)(\ln N)^2, \, k (\ln N)^2]$ is bounded by $(\ln N)^3$.

Once again, the essential contribution to the integral $\int_{tv}^\infty \cdots \, \mathrm{d} y$ on the right-hand side comes from the neighbourhood of $y=vt$; we write 
$$
\E(\# \widetilde{E}_t^{\mathrm{NBBM}})
\gtrsim
e^{-(\frac{v^2}{2\sigma^2} -1)t}\, 
\E \Big( {\bf 1}_{A_t} \, (\prod_{k=1}^{t/(\ln N)^2} {\bf 1}_{G_k})
\prod_{j: \, \tau_j \le t} {\bf 1}_{D_t^{\mathrm{NBBM}}(\tau_j)} \, \Big| \, X_t = vt\Big) \, .
$$

\noindent Compared to the discussions for the $L$-BBM in the previous subsection, we have a new factor $\prod_{k=1}^{t/(\ln N)^2} {\bf 1}_{G_k}$; conditionally on the path of $X$, the probability of $\cap_{k=1}^{t/(\ln N)^2} G_k$ is at least $(1-e^{-c_3 (\ln N)^2})^{t/(\ln N)^2}$ (for some constant $c_3>0$), which is greater than or equal to $e^{-t\, e^{-c_4 (\ln N)^2}}$ (for some constant $c_4>0$). As such, using again $\P_X$ to denote the conditional probability given $X$, we have
\begin{eqnarray*}
 && \E_X \Big( (\prod_{k=1}^{t/(\ln N)^2} {\bf 1}_{G_k})
\prod_{j: \, \tau_j \le t} {\bf 1}_{D_t^{\mathrm{NBBM}}(\tau_j)} \Big)
   \\
 &\ge&e^{-t\, e^{-c_4 (\ln N)^2}} \, \E_X \Big( \prod_{j: \, \tau_j \le t} \P_X(D_t^{\mathrm{NBBM}}(\tau_j) \, | \, \tau ) \, \Big| \, \bigcap_{k=1}^{t/(\ln N)^2} G_k \Big) .
\end{eqnarray*}

\noindent 
We have
$$
\E_X \Big( \prod_{j: \, \tau_j \le t} \P_X(D_t^{\mathrm{NBBM}}(\tau_j) \, | \, \tau ) \, \Big| \, \bigcap_{k=1}^{t/(\ln N)^2} G_k \Big)
\ge
\E_X \Big( \prod_{j: \, \tau_j \le t} \P_X(D_t^{\mathrm{NBBM}}(\tau_j) \, | \, \tau ) \Big) ,
$$

\noindent which equals $\exp \{ -2 \int_0^t [1- P_X(D_t^{\mathrm{NBBM}}(u))]\, \mathrm{d} u\}$. We can now carry out the same computations as in the case of the $L$-BBM, to see that
$$
\E(\# \widetilde{E}_t^{\mathrm{NBBM}}) 
\gtrsim
\exp \Big\{ -(\frac{v^2}{2\sigma^2} -1+e^{-c_4 (\ln N)^2})t - 3\int_0^t \E [1-\P_X (D_t^{\mathrm{NBBM}}(u))] \, \mathrm{d} u \Big\}\, .
$$

\noindent [This is the analogue for the $N$-BBM, of the inequality in \eqref{L-BBM:Jensen}.]

For all $s>0$ and $x\in (-\infty, \, \infty)$, let us write ${\cal N}(x,\, s)$ for the number of particles lying in $[x, \, \infty)$ at time $s$ in an BBM independent of $X$, and $M(s)$ the maximal position at time $s$ of the BBM. By definition of $D_t^{\mathrm{NBBM}}(u)$,
\begin{eqnarray*}
    1-\P_X (D_t^{\mathrm{NBBM}}(u)) 
 &\le& \P_X \Big( \exists s\in [(\ln N)^2, \, t-u]: \, M(s) \ge X_{s+u}-X_u \Big) +
    \\
 && + \P_X \Big( \exists s\in (0, \, t-u]: \, {\cal N}(X_{s+u}-X_u, \, s) \ge M \Big) \ .
\end{eqnarray*}

\noindent We argue that this implies
\begin{eqnarray*}
    1-\P_X (D_t^{\mathrm{NBBM}}(u)) 
 &\le& \int_{(\ln N)^2}^{t-u} \P_X \Big( M(s) \ge X_{s+u}-X_u \Big) \, \mathrm{d} s 
    \\
 && + \int_0^{t-u} \P_X \Big( {\cal N}(X_{s+u}-X_u, \, s) \ge M \Big) \, \mathrm{d} s \ ,
\end{eqnarray*}

\noindent even though the rigorous meaning of the inequality should be formulated as in the paragraph following \eqref{premier_moment_L-BBM:estimation_technique}.

The first probability expression on the right-hand side  
$\P_X  ( M(s) \ge X_{s+u}-X_u )$ is bounded by $\min [ 1, e^{s - \frac{(X_{s+u}-X_u)^2}{2\sigma^2s}} ]$. 
The probability $\P_X ( {\cal N}(X_{s+u}-X_u, \, s) \ge M )$ was denoted by $Q(X_{s+u}-X_u, \, s)$ in Section \ref{subs:N-BBM} (with $M$ in place of $N$), and we have seen in \eqref{q-NBBM} that 
$$
\P_X ( {\cal N}(X_{s+u}-X_u, \, s) \ge M )
\le
\min\Big[ 1, e^{s-\ln M  - \frac{(X_{s+u}-X_u)^2}{2\sigma^2s}}\Big] \, .
$$

\noindent As such,
\begin{eqnarray*}
    1-\P_X (D_t^{\mathrm{NBBM}}(u))
 &\le& \int_{(\ln N)^2}^{t-u} \Big[ {\bf 1}_{\{ X_{s+u}-X_u < (2\sigma^2 s^2)^{1/2} \} }
    \\
 &&\hskip-40pt +{\bf 1}_{\{ X_{s+u}-X_u \ge (2\sigma^2 s^2)^{1/2} \} } \exp\Big( s - \frac{(X_{s+u}-X_u)^2}{2\sigma^2 s} \Big) \Big] \, \mathrm{d} u
    \\
 &&+ \int_{\ln M}^{t-u} \Big[ {\bf 1}_{\{ X_{s+u}-X_u < [(2\sigma^2 s)(s-\ln M)]^{1/2} \} }
    \\
 &&\hskip-40pt +{\bf 1}_{\{ X_{s+u}-X_u \ge [(2\sigma^2 s)(s-\ln M)]^{1/2} \} } \exp\Big( s - \frac{(X_{s+u}-X_u)^2}{2\sigma^2 s} \Big) \Big] \, \mathrm{d} u \, .
\end{eqnarray*}

\noindent With the notation $X_u = vu + \sigma(W_u - \frac{u}{t}W_t)$ (where $W$ denotes again a standard Brownian motion with variance $1$, we have $X_{s+u}-X_u = vs + \sigma (W_{s+u}-W_u -\frac{s}{t} W_t)$. 
%
The random variable $W_{s+u}-W_u -\frac{s}{t} W_t$ has the Gaussian ${\cal N}(0, \, s(1-\frac{s}{t}))$ law. As for the $L$-BBM, some elementary computations yield that, in case $v>(4\sigma^2)^{1/2}$, the subtrees move forward faster than the usual speed $(2\sigma^2)^{1/2}$, whereas if $(2\sigma^2)^{1/2} <v \le (4\sigma^2)^{1/2}$, these subtrees make no particular effort, and wait only for the occasions when the red particle makes some fluctuations toward the left.\footnote{As we shall point out in Section \ref{s:conclusion}, this picture is probably inaccurate, and is only due to the fact that our upper bound for $Q(x, \, s)$ is not optimal. We conjecture that regardless of the value of $v$, the subtrees never make any particular effort in the $N$-BBM, which would be in complete contrast with the $L$-BBM.} Letting $t\to \infty$ and then $N\to \infty$, we obtain:
$$
\E(\# \widetilde{E}_t^{\mathrm{NBBM}})
\gtrsim
\exp\left[ -(\frac{v^2}{2\sigma^2} -1+e^{-c_4 (\ln N)^2} + M^{-\beta(v)+o_N(1)})t \right] \, ,
$$

\noindent where $\beta(v)$ is defined in \eqref{Res1-NBBM}, and $o_N(1)$ stands for a term not depending on $t$ such that $\lim_{N\to \infty} o_N(1) =0$. Note that $e^{-c_4 (\ln N)^2}$ is negligible compared to $M^{-\beta(v)+o_N(1)}$. This yields the desired lower bound \eqref{N-BBM:E(Et):lb}.

\subsection{First moment computations for the CBRW}

The proof for the CBRW is along the lines of the proof for the $L$-BBM and for the $N$-BBM. Let
$$
\widetilde{E}_t^{\mathrm{CBRW}}
:=
\bigcup_{i=1}^{{\cal N}(t)}\, \{ \hbox{\rm particle $i$ lies in $[vt, \, \infty)$, leans to the left, does not coalesce} \} \, .
$$

\noindent The meaning of "leans to the left" is as before: the path of the particle lies in $(-\infty, \, t'v+t^{2/3}]$ for all $t'\in [0, \, t]$. By "does not coalesce", we mean that at no time during $[0, \, t]$ does the particle coalesce with any other particle.

Let $\# \widetilde{E}_t^{\mathrm{CBRW}}$ denote the number of $i$ satisfying the conditions in $\widetilde{E}_t^{\mathrm{CBRW}}$. The main estimates for the CBRW are: 
\begin{eqnarray}
    \E(\# \widetilde{E}_t^{\mathrm{CBRW}})
 &\gtrsim& \exp\left[ -(f(v)-r+ \mu^{\gamma(v)+o_\mu(1)})t \right] \, ,
    \label{CBRW:E(Et):lb}
    \\
    \E[(\# \widetilde{E}_t^{\mathrm{CBRW}})^2]
 &\lesssim& \exp\left[ -(f(v)-r)t\right] \, ,
    \label{CBRW:E(Et):ub}
\end{eqnarray}

\noindent where $\gamma(v)$ and $f(v)$ are defined in \eqref{Res1-CBRW} and \eqref{parametric} respectively, and $o_\mu(1)$ stands for a term not depending on $t$ such that $\lim_{\mu\to 0} o_\mu(1) =0$. Equations \eqref{CBRW:E(Et):lb} and \eqref{CBRW:E(Et):ub} together with the Cauchy--Schwarz inequality will yield the upper bound stated in \eqref{Res-CBRW} for the large deviation function for the CBRW.

The proof of (\ref{CBRW:E(Et):ub}), which is identical for all the three models, is postponed to Subsection \ref{2nd_moment}. The rest of this subsection is devoted to the proof of \eqref{CBRW:E(Et):lb}. 

Writing $X=(X_u, \, u\in [0, \, t])$ again for the trajectory of the red particle $i$, and $A_t := \{ X_u \le uv+t^{2/3}, \, \forall u\in [0, \, t]\}$ as in \eqref{At}, we have 
$$
\E(\# \widetilde{E}_t^{\mathrm{CBRW}})
=
\sum_{k: \, vt \le k \sigma \le vt + t^{2/3}} e^{rt} \, P(k\sigma ; \, t) \, 
\E \Big( {\bf 1}_{A_t} \, \prod_{j: \, \tau_j \le t} {\bf 1}_{D_t^{\mathrm{CBRW}}(\tau_j)} \, \Big| \, X_t = k \sigma\Big) \, ,
$$

\noindent where, $P(k\sigma ; \, t)$ is the probability that a random walk is at position $k\sigma$ at time $t$, and for all $u\in [0, \, t]$, $D_t^{\mathrm{CBRW}}(u)$ stands for the event that none of the particles in the subtree of BBM branched at time $u$ on the path of $X$ coalesces with the red particle. Here, $(\tau_j, \, j\ge 1)$ is as before the atoms of a rate-2 Poisson process.

For $t\to \infty$, $P(k\sigma ; \, t) \sim e^{-tf(k\sigma/t)}$ where $f$ is as in (\ref{fv0}), and the essential contribution to the sum on the right-hand side comes from $k \approx \frac{vt}{\sigma}$; we treat $\frac{vt}{\sigma}$ as an integer, and write 
$$
\E(\# \widetilde{E}_t^{\mathrm{CBRW}})
\gtrsim
e^{t(r- f(v))} \, 
\E \Big( {\bf 1}_{A_t} \, \prod_{j: \, \tau_j \le t} {\bf 1}_{D_t^{\mathrm{CBRW}}(\tau_j)} \, \Big| \, X_t = vt\Big) \, .
$$

\noindent The same computations as for the $L$-BBM (see \eqref{L-BBM:Jensen}) give that
$$
\E(\# \widetilde{E}_t^{\mathrm{CBRW}})
\gtrsim
\exp \Big\{ t(r- f(v)) - 3\int_0^t \E[1- \P_X (D_t^{\mathrm{CBRW}}(u))] \, \mathrm{d} u \Big\}\, .
$$

\noindent As for the $L$-BBM, we argue that
$$
1-\P_X (D_t^{\mathrm{CBRW}}(u))
\le
\int_0^{t-u} \P_X (B_{u,s}) \, \mathrm{d} s \, ,
$$

\noindent where $B_{u,s}$ denotes the event that there exists a particle branched at time $u$ that coalesces with the red particle at time $u+s$. [For a rigorous meaning of this inequality, see the paragraph following \eqref{premier_moment_L-BBM:estimation_technique}.] By the Markov inequality, $\P_X(B_{u,s})$ is bounded by the $\P_X$-expected number of particles branched at time $u$ that coalesce with the red particle at time $u+s$, and this $\P_X$-expected number is approximately $\mu \, \exp[s(r- f( \frac{X_{u+s} - X_u}{s}))]$. On the other hand, $\P_X(B_{u,s})\le 1$. So 
$$
\P_X(B_{u,s})
\le
\min \Big[ 1, \, \mu \, \exp[s(r- f( \frac{X_{u+s} - X_u}{s}))] \Big] \, .
$$

\noindent Taking expectation with respect to the law of the red particle, we arrive that
\begin{eqnarray*}
    \E(\# \widetilde{E}_t^{\mathrm{CBRW}})
 &\gtrsim& \exp \Big\{ t(r- f(v))
    \\
 && \quad - 3\int_0^t \mathrm{d} u \, \int_0^{t-u} \mathrm{d} s \, \E\min \Big[ 1, \, \mu \, \exp[s(r- f( \frac{X_{u+s} - X_u}{s}))] \Big] \Big\}\, .
\end{eqnarray*}

\noindent {F}rom here, we can use the computations presented at the end of Section \ref{subs:heuristic_CBRW} (those leading to \eqref{constraint} and \eqref{v1}). This yields \eqref{CBRW:E(Et):lb}.

\subsection{Second moment computations for the three models}
\label{2nd_moment}

We use a common proof for \eqref{L-BBM:E(Et):ub} and \eqref{N-BBM:E(Et):ub}, for the $L$-BBM and the $N$-BBM, respectively. The proof of \eqref{CBRW:E(Et):ub}, for the CBRW, is along similar lines, and is omitted. 

It suffices to prove that
$$
\E(\Lambda_t^2)
\lesssim
e^{-(\frac{v^2}{2\sigma^2}-1)t} \ ,
$$

\noindent if $\Lambda_t = \Lambda_t(v)$ denotes the number of particles in the BBM (without selection) at time $t$ lying in $[vt, \, \infty)$ and leaning on the left (i.e., whose trajectories are in $(-\infty, \, vt'+t^{2/3}]$ for all $t'\in [0, \, t]$).

By definition,
$$
\E(\Lambda_t^2)
\le
\E(\Lambda_t)
+
\int_0^t \mathrm{d} \tau \int_{-\infty}^{v\tau +t^{2/3}} \mathrm{d} y \,  \frac{e^{\tau - \frac{y^2}{2\sigma^2 \tau}} }{(2\pi \sigma^2 \tau)^{1/2}} \Big( \int_{vt}^\infty \mathrm{d} z \, \frac{e^{(t-\tau) - \frac{(z-y)^2}{2\sigma^2 (t-\tau)}}}{(2\pi \sigma^2 (t-\tau))^{1/2}} \Big)^2 \ .
$$

\noindent [It is an inequality because the trajectories are not required to lean on the left, but only lie in $(-\infty, \, v\tau +t^{2/3}]$ at time $\tau$, when they split.] We have $\E(\Lambda_t) \le e^{-(\frac{v^2}{2\sigma^2}-1)t}$. 

It is convenient to split $\int_{-\infty}^{v\tau +t^{2/3}} \mathrm{d} y$ into the sum of $\int_{-\infty}^{v\tau} \mathrm{d} y$ and $\int_{v\tau}^{v\tau +t^{2/3}} \mathrm{d} y$.

Since $y\mapsto \frac{y^2}{2\sigma^2 \tau} + \frac{(z_1-y)^2}{2\sigma^2 (t-\tau)} + \frac{(z_2-y)^2}{2\sigma^2 (t-\tau)}$ is non-decreasing on $[0, \, v\tau]$ (for all $z_1 \ge vt$ and $z_2\ge vt$), it follows for the first integral that
\begin{eqnarray*}
 &&\int_0^t \mathrm{d} \tau \int_{-\infty}^{v\tau} \mathrm{d} y \,  \frac{e^{\tau - \frac{y^2}{2\sigma^2 \tau}} }{(2\pi \sigma^2 \tau)^{1/2}} \Big( \int_{vt}^\infty \mathrm{d} z \, \frac{e^{(t-\tau) - \frac{(z-y)^2}{2\sigma^2 (t-\tau)}}}{(2\pi \sigma^2 (t-\tau))^{1/2}} \Big)^2
    \\
 &\le&\int_0^t \mathrm{d} \tau \, e^{\tau - \frac{v^2 \tau}{2\sigma^2}} \Big( e^{(t-\tau) - \frac{v^2(t-\tau)}{2\sigma^2}} \Big)^2
    \\
 &\le& \int_0^t \mathrm{d} \tau \, e^{-(\frac{v^2}{2\sigma^2}-1)(2t-\tau)}
    \\
 &\lesssim& e^{-(\frac{v^2}{2\sigma^2}-1)t} ,
\end{eqnarray*}

\noindent using again our notation $a(t) \lesssim b(t)$ meaning that $\limsup_{t\to \infty} \frac{\ln [a(t)/b(t)]}{\ln t} \le 0$.

A few more lines of elementary computations show that the extra integral $\int_{v\tau}^{v\tau +t^{2/3}} \mathrm{d} y$ leads to an upper bound $e^{-(\frac{v^2}{2\sigma^2}-1)t + o(t)}$. Therefore, we get the claimed upper bound for $\E(\Lambda_t^2)$.\hfill$\Box$

\section{Conclusion}

\label{s:conclusion}

In the present paper we have  tried to estimate the large deviation function for the position of the rightmost particle of   three generalizations of the branching Brownian motion and of the branching random walks, subjected to selection or coalescence mechanisms.
We have proved the existence of  a large deviation function (\ref{ld-LBBM},\ref{ld-NBBM},\ref{ld-CBRW}) for positive deviations of the position
of the rightmost particle of these three models:
the $L$-BBM, the $N$-BBM and the CBRW.
For large $L$ for the $L$-BBM, large $N$  for the $N$-BBM, and for  small $\mu$ for the CBRW, we obtain  upper bounds for these large deviation functions (\ref{Res-LBBM},\ref{Res2-LBBM}), (\ref{Res-NBBM},\ref{Res1-NBBM}), (\ref{Res-CBRW},\ref{Res1-CBRW}).
Our results are limited to velocities larger than the typical velocity $v_c$  of the rightmost particle of the BBM or of the BRW.
Our approach does not allow us to give lower bounds for these large deviation functions.

It has been  shown by  duality  \cite{DMS} that the  coalescence branching random walk is closely related to the noisy version of the F-KPP equation. Of course it would be interesting to see whether a direct analysis of the noisy F-KPP equation could confirm our result (\ref{ld-CBRW},\ref{Res-CBRW},\ref{1.22}).

As recalled in the introduction, the F-KPP equation  gives the evolution the the probability distribution of the position of the rightmost particle of a BBM and the large deviation function of this position is (\ref{ld-BBM},\ref{psi-BBM}). One question we tried (without success) to solve and that  we would like to raise in this conclusion is how to obtain the probability $Q_N(x,t)$ of finding $N$ particles on the right of position $x=v t$   for $N \sim \ln t$. We 
could only get the following  lower bound 
\begin{equation}
Q_N(x,t) \gtrsim \min\left[ 1, e^{t - {x^2 \over 2 \sigma^2 (t-\ln  N)}}\right]
\label{conjecture}
\end{equation}
by considering the events where   a single  particle  moves first a distance $y$  during a time $s$ and then gives rise to a regular tree which produces $N$ particles on the right of $x$, 
i.e. 
$$
Q_N(x,t) \gtrsim \min\left[ 1, \max_{y,s} \left[e^{s - {y^2 \over 2 \sigma^2 s}}\right]
 \right] 
$$
where $y$ and $s$ are related by $$\ln N =(t- s)- { (x-y)^2 \over 2\sigma^2(t-s)} \ . $$

\noindent If the rhs of (\ref{conjecture}) were the true estimate and not simply a lower bound, $\beta(v)$ in the first line of (\ref{Res1-NBBM}) would remain valid even for $v>\sqrt{2}\, v_c\ $. 

The question of negative large deviations of the position of the rightmost particle (as considered in 
\cite{MS,MVS} for the CBRW) would also be interesting to attack.
In this case the result might   strongly depend on whether one starts with a single particle or more than one particle (in \cite{MS,MVS} it was assumed that the initial number  of particles is large and even infinite).
If one starts with a single particle, one would get already  for the large deviation  function 
(\ref{ld-BBM}) of the BBM
\begin{equation}
 \psi_{\rm BBM}(v) = \begin{cases}
 2(\sqrt{2} -1) \left( 1- {v\over v_c} \right)  & {\rm for }  \ \ \  -(\sqrt{2}-1) v_c < v < v_c
\\ & \\
  1+ {v^2  \over  v_c^2}  & {\rm for }  \ \ \  v < -(\sqrt{2}-1) v_c 
   \end{cases}
\label{Res1bis-NBBM} \ ,
\end{equation}
(where   $v_c=\sqrt{2} \sigma$)
and the events which would dominate the contributions to the large deviations of the $L$-BBM, $N$-BBM and the CBRW would be rather different from those considered in the present paper.

\bigskip
\bigskip

B.D.\ thanks the LPMA in Jussieu for its hospitality for the whole academic year 2014--2015.

\end{document}